\documentclass{ws-procs9x6}
\def\beq{\begin{equation}}
\def\eeq{\end{equation}}

\def\na{\nabla}
\def\pa{\partial}

\def\al{\alpha}
\def\be{\beta}

\def\ga{\gamma}

\def\vp{\varepsilon}

\def\la{\lambda}

\def\si{\sigma}

\def\ph{\varphi}

\def\Ga{\Gamma}

\begin{document}

\title{
EFFECTIVE QFT AND WHAT IT TELLS US ABOUT DYNAMICAL TORSION
} 

\author{ILYA L.\ SHAPIRO}

\address{Departamento de F\'{i}sica,
Universidade Federal de Juiz de Fora
\\
36036-330, Juiz de Fora, Minas Gerais, Brazil\\
E-mail: shapiro@fisica.ufjf.br}

\begin{abstract}
The covariantly constant spacetime torsion is one of the fields
which may break Lorentz and CPT symmetry. We review the previous
works on the dynamical torsion in the framework of effective
quantum field theory (QFT). It turns out that the existence of
propagating torsion is strongly restricted by the QFT principles.
In particular, the torsion mass must be much greater than the
masses of all fermionic particles. In this situation, the main
chance to observe torsion is due to some symmetry breaking which
may, in principle, produce almost constant background torsion
field.
\end{abstract}

\bodymatter

\section{Introductory note}

In this contribution to the CPT'10 Proceedings
I decided to review our papers devoted to the
effective QFT approach to the problem of dynamical
torsion. The content is not original and is based on
Refs.\ \refcite{betor,guhesh} (see also Ref.\ \refcite{torsi}), but the
purpose is to present it in a maximally simple and
qualitative form.

\section{Classical torsion gravity and renormalization}

An extended version of QED with the general Lorentz and CPT
symmetry breaking terms is \cite{CalKos-97}
\beq
S = \int d^4x\sqrt{-g}\,\left\{\,
\frac{i}{2}\,\bar{\psi}\Ga^\mu D_\mu\psi
- \frac{i}{2}\, D^{\star}_\mu\bar{\psi}\Ga^\mu\psi
-\bar{\psi}\, M\,\psi -\frac{1}{4} F_{\mu\nu}\,F^{\mu\nu}
\right.
\nonumber
\eeq
\beq
-  \left.
\frac{1}{4}\, (k_F)_{\mu\nu\al\be}\,F^{\mu\nu}\,F^{\al\be}
+ \frac{1}{2}\,(k_{AF})^\rho \,\vp_{\rho\la\mu\nu}
\,A^\la\,F^{\mu\nu}\,\right\}\,.
\label{1}
\eeq
Here
\beq
D_\mu = \na_\mu + i\,q\,A_\mu\, ; \quad
\Ga^\nu = \ga^\nu + \Ga_1^\nu\, ; \quad
M = m + M_1\,,
\label{2}
\eeq
and the quantities $\Ga_1^\nu$ and $M_1$ are given by
\beq
\Ga_1^\nu  =  c^{\mu\nu}\ga_\mu + d^{\mu\nu} \ga_5\ga_\mu
+ e^\nu + i\,f^\nu\ga_5
+\frac{1}{2}\,g^{\la\mu\nu}\si_{\la\mu}\, ,
\nonumber
\eeq
\beq
M_1  =  a_\mu\,\ga^\mu + b_\mu\,\ga_5\,\ga^\mu
+ i\, m_5\ga_5 + \frac{1}{2}\, H_{\mu\nu}\,\si^{\mu\nu}\, .
\label{3}
\eeq
One of the most phenomenologically important terms is
the one which includes $b_\mu=\eta S_\mu$, where $S_\mu$
is dual to the spacetime torsion.

In a spacetime with independent metric $g_{\mu\nu}$
and torsion $T^\alpha_{\,\beta\gamma}$ the connection is
nonsymmetric,
$\tilde{\Gamma}^\alpha_{\,\beta\gamma} -
\tilde{\Gamma}^\alpha_{\,\gamma\beta} =
T^\alpha_{\,\beta\gamma}$. \
Using the metricity condition
$\tilde{\nabla}_\mu g_{\alpha\beta} = 0$ we get
\beq
\tilde{\Gamma}^\alpha_{\,\beta\gamma}
= {\Gamma}^\alpha_{\,\beta\gamma} +
K^\alpha_{\,\beta\gamma}\,,\quad
{\Gamma}^\alpha_{\,\beta\gamma}
= \left\{\,^\alpha_{\,\beta\gamma} \right\}
\,,\quad
K^\al_{\,\be\ga} = \frac{1}{2} \left( T^\al_{\,\,\be\ga} -
T^{\,\alpha}_{\beta\,\gamma} - T^{\,\alpha}_{\gamma\,\beta} \right).
\label{4}
\eeq
It is convenient to divide torsion into irreducible components
\beq
T_{\beta} = T^\alpha_{\,\beta\alpha}
,\,\,
S^{\nu} = \varepsilon^{\alpha\beta\mu\nu}T_{\alpha\beta\mu}
,\,\,
q^\alpha_{\,\beta\gamma}
,\,\,\,
\mbox{where}
\,\,\,
q^\alpha_{\,\beta\alpha} =
\varepsilon^{\alpha\beta\mu\nu}q_{\alpha\beta\mu} = 0\,.
\label{5}
\eeq
The action of a spinor field minimally coupled to torsion is
\beq
S = \int d^4 x \,e \,
\Big\{ \frac{i}{2}\bar{\psi}\gamma^\mu \tilde{\nabla}
_\mu \psi - \frac{i}{2}\tilde{\nabla}_\mu\bar{\psi}\gamma^\mu\psi
+ m\bar{\psi}\psi \Big\}           
\nonumber
\eeq
\beq
= \int d^4 x\,e\,\,
\Big\{i\bar{\psi}\gamma^\mu(\nabla_\mu+\frac{i}{8}\gamma_5
S_\mu)\psi+m\bar{\psi}\psi\Big\}\,,
\label{6}
\eeq
where $\nabla_\mu$
is the Riemannian covariant derivative (without torsion).

Let us briefly consider renormalization of a gauge QFT in a
spacetime with torsion \cite{bush-85}.
In order to achieve consistent renormalizable theory, one
has to introduce interaction of matter fields
with metric and torsion, plus a vacuum action.
With scalars $\,\ph\,$ torsion interacts only nonminimally,
\beq
S_{sc} = \int d^4x\,\sqrt{-g}\,\,
\Big(\frac12\,g^{\mu\nu}\,\pa_\mu\ph\,\pa_\nu\ph
+\frac12\,m^2\ph^2 + \frac12\,\xi_i P_i\ph^2 \Big)\,,
\label{7}
\eeq
where $\,\xi_i\,$ are nonminimal parameters and
the relevant invariants are
\beq
P_1 = R
,\quad
P_2 = \na_\al T^\al
,\quad
P_3 = T_\al^2
,\quad
P_4 = S_\al^2
,\quad
P_3 = q_{\al\be\ga}^2\,.
\label{8}
\eeq
We assume that gauge vector fields do not interact with
torsion, since such interaction, generally, contradicts gauge
invariance.
On top of this we need a vacuum
(metric and torsion-dependent) action, which is, in general, rather
complicated \cite{chris}.
As far as we included all torsion-dependent terms which
can be met in the counterterms, the theory described above
is renormalizable.

It is important that the spinor field interacts with torsion
nonminimally,
\beq
S = \int d^4 x\,\sqrt{-g}
\,\big\{i\bar{\psi}\gamma^\mu(\na_\mu
+i\eta\ga_5 S_\mu + i\eta_2T_\mu)\psi + m\bar{\psi}\psi\big\}\,.       
\label{9}
\eeq
The minimal interaction corresponds to
$\eta = 1/8$, $\eta_2 = 0$.

The  $\be$-function for the nonminimal parameter $\eta$
has universal form \cite{bush-85,book},
\ $\be_\eta = Ch^2\eta$\,,
where $h$ is the Yukawa coupling and the model-dependent coefficient
$C$ is always positive. It is easy to see that interaction
with background torsion gets stronger in the UV. Moreover,
although torsion is a geometric field, the parameters $\eta$ have to
be different for distinct fermions, because within the Standard
Model their Yukawa couplings have different values.

\section{Dynamical torsion and effective approach}

What is the simplest possible action describing dynamical torsion?
For the sake of simplicity, consider antisymmetric torsion
and a flat metric. Let us try to construct
the effective QFT for dynamical (propagating) torsion.
The consistency conditions include unitarity of the $S$-matrix
and gauge-invariant but not power-counting renormalizability.
A particularly important aspect is that we can neglect possible
higher derivative terms. Application of the same approach to
quantum gravity \cite{don-94} led to some interesting results,
although its technical realization remains doubtful \cite{Polemic}.

Let us start from the action of the fermion field
\beq
S_{1/2}\,=\, i \int d^4x\,{\bar \psi}\, \left[
\,\ga^\al \,\left( {\pa}_\al - ieA_\al + i\,\eta\,\ga_5\,S_\al\,\right)
- im \,\right]\,\psi .
\label{10}
\eeq
There are two gauge symmetries, and the second one is softly broken:
\beq
\psi' = \psi\,e^{\al(x)}
,\,\,\,\,\,\,\,\,
{\bar {\psi}}' = {\bar {\psi}}\,e^{- \al(x)}
,\,\,\,\,\,\,\,\,
A_\mu ' = A_\mu - {e}^{-1}\, \pa_\mu\al(x)\,;
\nonumber
\eeq
\beq
\psi' = \psi\,e^{\ga_5\be(x)}
,\,\,\,\,\,\,
{\bar {\psi}}' = {\bar {\psi}}\,e^{\ga_5\be(x)}
,\,\,\,\,\,\,
S_\mu ' = S_\mu - {\eta}^{-1}\, \pa_\mu\be(x)\,.
\label{11}
\eeq
These symmetries lead to the form of the torsion action
\beq
S_{tor} = \int d^4x\,\left\{\, -a
\,S_{\mu\nu}S^{\mu\nu} + b\,(\pa_\mu S^\mu)^2
+ M_{ts}^2\,S_\mu S^\mu\,\right\}\,,
\label{12}
\eeq
where $\,S_{\mu\nu} = \pa_\mu S_\nu - \pa_\mu S_\nu$.
In the unitary theory the longitudinal and transverse
modes can not propagate simultaneously. Hence, one has to
choose one of the parameters $\,a,b\,$ to be zero. Indeed,
the only correct choice is $b=0$, because (\ref{11}) holds
in the renormalization of the massless sector.

Thus the torsion action is given by \cite{betor}
\beq
S_{tor} = \int d^4 x\,\left\{\, -\frac14\,S_{\mu\nu}S^{\mu\nu}
+ M_{ts}^2\, S_\mu S^\mu\,\right\}\,.
\label{13}
\eeq
The phenomenological analysis shows the torsion parameters
$M_{ts}$ and $\eta$ can be strongly restricted.
\cite{npt}$^,$\footnote{Earlier, the restrictions for the
parameters of purely longitudinal axial torsion vector
were established in Ref.\ \refcite{carroll}.}
However, torsion still can be used, e.g., as an alternative to
technicolor models \cite{zubkov-10}.

The most important question is whether it is really possible
to preserve unitarity at the quantum level. The answer to this
question is negative. Even in QED, without scalar fields, the
one-loop calculations show that
$\big({\bar \psi}\psi\big)^2$-type counterterms are necessary
at the one-loop level \cite{guhesh,CPT-Buch-06}. Then, in the
second loop order we meet a longitudinal divergence for the
torsion axial vector, $\int d^4x \,\big(\pa_\al S^\al\big)^2$.

The violation of unitarity in the fermionic sector occurs
when the following condition is not fulfilled \cite{guhesh}:
\beq
\eta\,\frac{m_{fermion}}{M_{ts}} \ll 1\,.
\label{15}
\eeq
As far as this constraint must hold for all fermions of the
MSM, the torsion mass must of the order of, at least, a few
TeV, or the coupling $\eta$ should be extremely small.
In both cases, there is no real chance to observe propagating
torsion.

\section{Conclusions}

We have shown, as a result of some complicated calculations in
Ref.\ \refcite{npt} and especially in Ref.\ \refcite{guhesh}, that 
the existence of the propagating torsion does contradict QFT
principles, even in the weakest possible effective version.
This result can be seen from two different viewpoints. First,
it means that we can restrict the spacetime geometry
using effective QFT arguments. As far as we know, this is
a unique known example of this sort. On the other hand, the
unique `chance of survival' for torsion is to show up as a
background field $b_\mu=\eta S_\mu$. The possible mechanisms
for the background torsion from symmetry breaking in gravity
have been discussed in Ref.\ \refcite{torsi} and Ref.\ 
\refcite{Kost-04}. So,
the search of torsion is essentially equivalent to looking for
a $b_\mu$ term in QED (\ref{1}) and its extensions in other
sectors of the Standard Model. An interesting discussion of 
possible extensions and last observational constraints has 
been given in Ref.\ \refcite{Kost-08}.

\section*{Acknowledgments}
I am very grateful to A. Belyaev, J. Helayel-Neto and G. de
Berredo-Peixoto for numerous discussions of the issue and
collaboration. The work of the author was partially supported
by CNPq, FAPEMIG, FAPES and ICTP. The special support of
FAPEMIG and FAPES for taking part in the CPT'10 Meeting is
gratefully acknowledged.


\end{document}